\documentclass[aps,pre,showpacs,twocolumn,floatfix]{revtex4}

\usepackage{amsmath}
\usepackage{epsfig}
\usepackage{mathrsfs}
\usepackage{psfrag}
\usepackage[T1]{fontenc}

\begin{document}

\title{Statistical Effects in the Multistream Model for Quantum
  Plasmas} 

\author{Dan Anderson, Bj\"orn Hall, Mietek Lisak and Mattias Marklund}   

\affiliation{Department of Electromagnetics,
  Chalmers University of Technology, SE--412 96 G\"oteborg, Sweden}

\date{\today}

\begin{abstract} 
A statistical multistream description of quantum plasmas is formulated,
using the Wigner--Poisson system as dynamical equations. 
A linear stability analysis of this system is carried out, and it is shown
that a Landau-like damping of plane wave perturbations occurs due to
the broadening of the  
background Wigner function that arises as a consequence of statistical
variations of the wave function phase. The Landau-like damping is
shown to suppress instabilities of the one- and two-stream type.
\end{abstract}


\maketitle


 \section{Introduction}

It has recently been pointed out \cite{Haas-Manfredi-Feix} that the
persistent trend towards increased miniaturization of electronic
devices implies that quantum effects will become important also for
certain transport processes, for which so far classical models have
been sufficient. An example of such a generalized transport equation,
in the form the Schr\"odinger--Poisson equation was analyzed in Ref.\
\cite{Haas-Manfredi-Feix}.
This analysis of a quantum plasma is based on the hydrodynamic
formulation of the 
Schr\"odinger--Poisson system, where macroscopic plasma quantities
such as density and average velocity are introduced. However, the
analysis does not take into account statistical (or kinetic) effects
associated with the finite width of the probability distribution
function. Kinetic effects are well-known in plasma physics, where
they may lead to the phenomenon of Landau damping.

The possibilities of using a general approach based on the Wigner
function \cite{Wigner,Moyal} was commented upon in Ref.\
\cite{Haas-Manfredi-Feix}, but only a simpler
approach based on macroscopic quantities was used. 
Obviously, in doing so the
possibilities of Landau-damping like effects are lost. In fact, the
possibility of obtaining Landau damping is also mentioned in Ref.\
\cite{Haas-Manfredi-Feix}, although in connection with a possible
generalization to the multi-stream case, in accordance with the
classical picture of Dawson \cite{Dawson}.
Particular attention was given to the classical one- and two stream
instabilities in a cold plasma and it was shown that the main quantum
effect on the wave propagation could be characterized as a generalized
dispersion. 

However, recently much attention, within the nonlinear optics
community, has been devoted to effects of partial wave incoherence e.g.\
in the form of phase noise on a constant amplitude wave
\cite{Hall-etal,Christodoulides-etal,Mitchell-etal}. In particular, it
has been shown in Ref.\ \cite{Hall-etal}, where the Wigner transform
was introduced as a means to study the modulational instability of an
optical plane wave, that the phase noise gives
rise to a Landau-like damping effect on the one stream
modulational instability. 

It is the purpose of the present work to generalize 
the analysis made in Ref.\ \cite{Haas-Manfredi-Feix} by analyzing the
properties of the one- and two-stream instabilities in a quantum
plasma using the Wigner formalism and including the effect of phase
noise developed in Ref.\ \cite{Hall-etal}. The results clearly show the
suppressing effect on the instabilities due to the Landau-like damping
effect caused by the phase noise of the Wigner function.

\section{Quantum statistical dynamics}

In non-relativistic many-body problems, the Wigner transformation is a
useful means to derive equations describing the quantum statistical
dynamics of the system of interest. Thus, one is able to
generalize the classical Vlasov equation to a quantum mechanical
regime, in the sense that the dynamical equation for the Wigner
function describes particles moving in a self-consistent force field
and in such a way that  
the evolution equation for the Wigner function
takes the form of its classical analogue in the limit $\hbar
\rightarrow 0$. 

Haas et al.\ \cite{Haas-Manfredi-Feix} have considered the dynamics of a
quantum plasma described by the nonlinear Schr\"odinger--Poisson
system of equations:
\begin{subequations}\label{eq:Schrodinger-Poisson}
\begin{eqnarray}
  i\hbar\frac{\partial\psi_i}{\partial t} +
    \frac{\hbar^2}{2m}\frac{\partial^2\psi_i}{\partial x^2} +
    e\phi\psi_i = 0  \ , \label{eq:Schrodinger} \\
  \frac{\partial^2\phi}{\partial x^2} = \frac{e}{\varepsilon_0} 
    \left( \sum_{i = 1}^N \langle|\psi_i|^2\rangle - n_0 \right) \ ,
    \label{eq:Poisson} 
\end{eqnarray}
\end{subequations}
where $i = 1, ..., N$ numbers the electrons as described by pure
states, with $\psi_i$ being the wave function for each such state;
$\phi(x,t)$ is the electrostatic potential, while $m$ and $-e$ are the
mass and charge of the electrons, respectively. 
The fixed ion background has the density
$n_0$. Following Ref.\ \cite{Hall-etal}, 
we have introduced the Klimontovich statistical average, 
denoting it by $\langle\cdot\rangle$. 
The statistical averaging
becomes important when the wave function contains e.g.\ a 
stochastically varying phase \cite{Hall-etal}.

In Ref.\ \cite{Haas-Manfredi-Feix}, the one-stream and two-stream
models have been investigated 
and the dispersion relation for the two-stream instability was derived,
showing an appearance of a new, purely quantum branch. 
We note that the analysis presented in Ref.\ \cite{Haas-Manfredi-Feix}
is based on the hydrodynamic formulation of the system
(\ref{eq:Schrodinger-Poisson}), where macroscopic plasma
quantities, such as
density and average velocity, are introduced. However, this type of
analysis does not take into account statistical properties
of the wave function that may lead to a broadening of the
probability distribution function. In fact, such effects may give rise
to a Landau-like damping both in the case of the
single-stream and two-stream instabilities. 

In order to take the statistical effects into account, it is
convenient to introduce
the Wigner distribution function $W_i(x, t; p)$, corresponding to the
wave function $\psi_i(x, t)$, as
\begin{widetext}
\begin{equation}\label{eq:Wigner}
  W_i(x, t; p) = \frac{1}{2\pi\hbar} \int_{-\infty}^{+\infty} dy \,
  \exp(ipy/\hbar) \langle \psi_i^*(x + y/2, t)\psi_i(x - y/2, t) \rangle\ ,
\end{equation}
\end{widetext}
which has the property 
\begin{equation}
  \int_{-\infty}^{+\infty} dp \, W_i(x, t; p) = 
    \langle|\psi_i(x,t)|^2\rangle \ .
\end{equation}

Using Eq.\ (\ref{eq:Wigner}), Eq.\ (\ref{eq:Schrodinger}) can be
formulated as a kinetic equation for the Wigner distribution, viz the
Wigner--Moyal equation
\begin{equation}\label{eq:Wignereq}
  \left(\frac{\partial}{\partial t} +
  \frac{p}{m}\frac{\partial}{\partial x}\right)W_i +
  \frac{2e}{\hbar}\phi\sin\left( 
  \frac{\hbar}{2}\frac{\stackrel{\leftarrow}{\partial}}{\partial x}
  \frac{\stackrel{\rightarrow}{\partial}}{\partial p} \right)W_i = 0 \
  ,
\end{equation}
where the sine-operator is defined in terms of its Taylor expansion.
Correspondingly, Eq.\ (\ref{eq:Poisson}) can be rewritten as
\begin{equation}\label{eq:Poisson2}
  \frac{\partial^2\phi}{\partial x^2} = \frac{e}{\varepsilon_0}\left( 
    \sum_{i = 1}^{N}\int_{-\infty}^{+\infty} dp \, W_i -n_0 \right) \
    .
\end{equation}
Clearly, an equilibrium solution of Eqs.\ (\ref{eq:Wignereq}) and
(\ref{eq:Poisson}) is $\phi = 0$ and $W_i = W_{i0}(p)$.

In order to study the modulational stability of the system
(\ref{eq:Wignereq})--(\ref{eq:Poisson2}), we introduce a small
perturbation according to
\begin{subequations} 
\begin{eqnarray}
  W_i(x,t;p) = W_{i0}(p) + \widetilde{W}_i\exp[i(Kx - \Omega t)] \ ,
  \\
  \phi(x, t) = \tilde{\phi}\exp[i(Kx - \Omega t)] \ , 
\end{eqnarray}
\end{subequations}
where $|\widetilde{W}_i| \ll |W_{i0}|$ and $K$ and $\Omega$ are the
wave number and frequency of the perturbation, respectively. 
The fact that the background distribution $W_{i0}$ is assumed to be
only a function of $p$ corresponds to the assumption of a plane wave
function with constant amplitude, but with a stochastically varying
phase, the characteristic properties of which are expressed by
$W_{i0}(p)$. Linearizing Eqs.\
(\ref{eq:Wignereq}) and (\ref{eq:Poisson2}), we obtain 
\begin{subequations}\label{eq:Wigner-Poisson}
\begin{eqnarray}
  -i\left(\Omega - \frac{p}{m}K\right)\widetilde{W}_i +
   \frac{2e}{\hbar}\tilde{\phi} \sin\left( \frac{i\hbar K}{2}
   \frac{\stackrel{\rightarrow}{\partial}}{\partial p} \right) W_{i0}
   = 0 \ , \\
  -K^2\tilde{\phi} = \frac{e}{\varepsilon_0}\sum_{i = 1}^{N} 
    \int_{-\infty}^{+\infty} dp \, \widetilde{W}_i  \ ,
\end{eqnarray}
\end{subequations}
where $\tilde{\phi}$ is the potential perturbation. Note that the fact
that the unperturbed potential $\phi$ is $\phi_0 = 0$ means that 
\begin{equation}
  \sum_{i = 1}^{N} 
    \int_{-\infty}^{+\infty} dp \, W_{i0} = n_0 \ .
\end{equation}

Eliminating $\tilde{\phi}$ in Eqs.\ (\ref{eq:Wigner-Poisson}), we obtain
the dispersion relation
\begin{equation}\label{eq:disprel}
  \frac{2ie^2m}{\varepsilon_0\hbar K^3} \sum_{i = 1}^{N}
  \int_{-\infty}^{+\infty} dp \, \frac{1}{p - m\Omega/K}
   \sin\left( \frac{i\hbar K}{2}
   \frac{\stackrel{\rightarrow}{\partial}}{\partial p} \right)W_{i0} +
  1 = 0\ .
\end{equation}
Using the fact that 
\begin{eqnarray}
  && 2\sin\left(\frac{i\hbar K}{2}
   \frac{\stackrel{\rightarrow}{\partial}}{\partial p}
   \right)W_{i0}(p)  \nonumber \\
  && \quad\qquad = i\left[ W_{i0}(p + \hbar K/2) - W_{i0}(p - \hbar
   K/2) \right] \ ,
\end{eqnarray}
relation (\ref{eq:disprel}) can be written in the form
\begin{equation}\label{eq:disprel2}
  1 = \frac{e^2m}{\varepsilon_0\hbar K^3}\sum_{i = 1}^{N}
  \int_{-\infty}^{+\infty} dp \, \frac{W_{i0}(p + \hbar K/2) 
     - W_{i0}(p - \hbar K/2)}{p - \Omega m/K} \ .
\end{equation}
Note that the pole $p = \Omega m/K$ gives rise to both a principal
part and an imaginary residue, as in the classical analysis of Landau
damping in plasma physics. 

Let us now consider the cases of one-stream and two-stream plasmas.

\subsection{One-stream plasma}

The dispersion relation (\ref{eq:disprel2}) reduces to 
\begin{equation}\label{eq:onestreamdisp}
  1 = \frac{e^2m}{\varepsilon_0\hbar K^3}
  \int_{-\infty}^{+\infty} dp \, \frac{W_{0}(p + \hbar K/2) 
     - W_{0}(p - \hbar K/2)}{p - \Omega m/K} \ ,
\end{equation}
where $W_0 \equiv W_{10}$. For a one-component Wigner spectrum with a
deterministic phase, i.e.\ a monoenergetic beam, $W_0(p)$ is given by 
\begin{equation}
  W_0(p) = n_0 \delta(p - p_0) \ ,
\end{equation}
which corresponds to a monochromatic plane wave function with constant
amplitude and phase. Equation (\ref{eq:onestreamdisp}) then yields
\begin{equation}
  1 = \frac{n_0e^2m}{\varepsilon_0 K^2}\frac{1}{(p_0 - \Omega m/K)^2 -
  \hbar^2K^2/4} \ ,
\end{equation}
i.e., 
\begin{equation}\label{eq:onestreamfluid}
  (\Omega - v_0K)^2 = \omega_p^2 + \frac{\hbar^2K^4}{4m^2} \ ,
\end{equation}
where $v_0 \equiv p_0/m$ and $\omega_p^2 \equiv
n_0e^2/m\varepsilon_0$. The expression (\ref{eq:onestreamfluid}) is
exactly the same as the one obtained in Ref.\
\cite{Haas-Manfredi-Feix}. It shows that quantum effects give rise to
wave dispersion for short wave-lengths. 

Let us now assume that the phase $\varphi(x)$ of the wave function 
$\psi_0$ varies stochastically, and that the corresponding correlation function
is given by
\begin{equation}
  \langle e^{-i\varphi(x + y/2)}e^{i\varphi(x - y/2)} \rangle = 
  e^{-p_T|y|} \ .
\end{equation}
This corresponds to the Lorentzian spectrum
\begin{equation}
  W_0(p) = \frac{n_0}{\pi}\frac{p_T}{(p - p_0)^2 + p_T^2} \ , 
\end{equation}
and the dispersion relation (\ref{eq:onestreamdisp}) now yields
\begin{equation}
  \Omega - \frac{p_0}{m}K = \left( \omega_p^2 + \frac{\hbar^2K^4}{4m^2}
  \right)^{1/2} - i\frac{p_T}{m}K \
  ,
\end{equation}
This result implies a completely new effect, a Landau-like damping due
to the width of the spectral distribution describing the stochastic
variation of the phase, i.e.\ due to the partial incoherence of the
beam. Furthermore, the damping effect increases with increasing
incoherence, i.e.\ with increasing $p_T$.


\subsection{Two-stream plasma}

According to Eq.\
(\ref{eq:disprel2}), the dispersion relation becomes
\begin{eqnarray}
  1 &=& \frac{e^2m}{\varepsilon_0\hbar K^3}\int_{-\infty}^{+\infty} dp
    \, \left[ \frac{W_{10}(p + \hbar K/2)  - W_{10}(p - \hbar K/2)}{p -
    \Omega m/K} \right. \nonumber \\ 
    &&\qquad \left. + \frac{W_{20}(p + \hbar K/2)  - W_{20}(p - \hbar
    K/2)}{p - \Omega m/K} \right ] \ .
\label{eq:twostreamdisp}
\end{eqnarray} 
For monochromatic beams with 
\begin{equation}
  W_{j0}(p) = n_{0j} \delta(p - p_{0j}) \ ; \ j = 1, 2 \ , 
\end{equation}
we get from Eq.\ (\ref{eq:twostreamdisp}) 
\begin{eqnarray}
  1 &=& \frac{\omega_{p1}^2}{(\Omega - p_{01}K/m)^2 -
  \hbar^2K^4/4m^2} \nonumber \\ 
   &&\qquad  + \frac{\omega_{p2}^2}{(\Omega - p_{02}K/m)^2 -
  \hbar^2K^4/4m^2} \ ,
\label{eq:twostreamfluid} 
\end{eqnarray}
where $\omega_{pj}^2 = e^2n_{0j}/\varepsilon_0m$ and $n_{01} + n_{02}
= n_0$. If we follow Ref.\ \cite{Haas-Manfredi-Feix} and consider the
symmetric case where $n_{01}
= n_{02} = n_0/2$, $p_{01} = -p_{02} \equiv p_0$, we obtain 
\begin{eqnarray}
  && \bar{\Omega}^4 - \left( 1 + 2\bar{K}^2 + \frac{H^2\bar{K}^4}{2}
    \right)\bar{\Omega}^2 \nonumber \\ 
  &&\, - \bar{K}^2\left( 1 -
    \frac{H^2\bar{K}^2}{4}\right) \left( 1 - \bar{K}^2 +
    \frac{H^2\bar{K}^4}{4}\right)  = 0 
\label{eq:twostreamfluid2}
\end{eqnarray}  
from Eq.\ (\ref{eq:twostreamfluid}). Here we have introduced
dimensionless variables according to 
\begin{equation} \label{eq:dimlessvar}
  \bar{\Omega} = \Omega/\omega_{p0} \ , \quad \bar{K} =
  p_0K/\omega_{p0}m \ , \quad H = \hbar\omega_{p0}m/p_0^2 \ .  
\end{equation}
Equation (\ref{eq:twostreamfluid2}) is
identical to the result obtained from the hydrodynamical theory, as
in Ref.\ \cite{Haas-Manfredi-Feix}. The solution of Eq.\
(\ref{eq:twostreamfluid2}) is 
\begin{equation}\label{eq:constraint1}
  \bar{\Omega}^2 = \frac{1}{2} + \bar{K}^2 + \frac{H^2\bar{K}^4}{4}
  \pm \frac{1}{2}\sqrt{1 + 8\bar{K}^2 + 4H^2\bar{K}^6} \ , 
\end{equation}
which implies $\bar{\Omega}^2 < 0$ and concomitant instability if 
\begin{equation}\label{eq:constraint2}
  (H^2\bar{K}^2 - 4)(H^2\bar{K}^4 - 4\bar{K}^2 + 4) < 0 \ .
\end{equation} 
This condition can be written
\begin{equation}
  1 - \frac{1}{\bar{K}^2} < \frac{H^2\bar{K}^2}{4} < 1 \ , 
\end{equation}
which reduces to the well-known two-stream instability result $K^2 < 1$
in the classical limit $H \rightarrow 0$. 

\begin{figure}
\centering\scalebox{.5}{\includegraphics{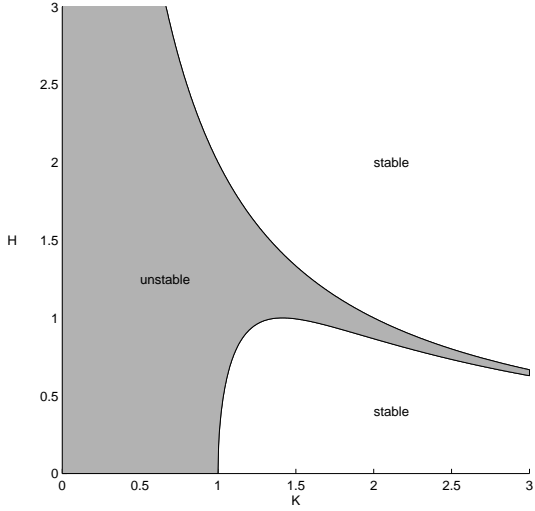}}
\caption{Qualitative plot of the stability/instability regions for the
two-stream quantum plasma, neglecting stochastic damping.}
\label{fig1}
\end{figure}

However, we infer from Eq.\ (\ref{eq:constraint2}) that the quantum
effect has a subtle influence on the instability. Equation
(\ref{eq:constraint2}) implies instability when the following condition
is satisfied in $(\bar{K}, H)$ space, viz
\begin{equation}
  H_-^2(\bar{K}) \equiv \frac{4}{\bar{K}^2}\left( 1 -
  \frac{1}{\bar{K}^2} \right) < H^2 
  < \frac{4}{\bar{K}^2} \equiv H_+^2(\bar{K}) \ .
\end{equation}
A qualitative plot of this is given in Fig.\ \ref{fig1} (a similar
figure and discussion was given in Ref.\ \cite{Haas-Manfredi-Feix},
but for later reference we present the figure and a discussion
related to it).  

Figure \ref{fig1} implies that when $H=0$, instability occurs only for $0 <
\bar{K} < 1$. However, when $H \neq 0$, a more complicated picture
emerges. In fact, as is seen from Fig.\ \ref{fig1}, the quantum effect plays
both a stabilizing and a destabilizing role. For $H > 1$, instability
occurs for all $\bar{K}$ such that $0 \leq \bar{K} \leq K_+(H) \equiv
2/H$. Thus, for $1 \leq H \leq 2$, the region of instability is
increased, whereas for $H \geq 2$ it is decreased as compared to the
case $H = 0$. 

For $H < 1$, instability occurs in two $K$-bands, viz $0 \leq \bar{K}
\leq K_-^{(1)}(H)$ and $K_-^{(2)}(H) \leq \bar{K} \leq K_+(H)$, where
$K_-^{(1,2)}(H)$ are the two solutions of the equation $1 -
1/\bar{K}^2 = H^2\bar{K}^2/4$, i.e.\
\begin{subequations}
\begin{eqnarray}
  K_-^{(1)}(H) = \frac{2}{H^2}\left(  1 + \sqrt{1 - H^2} \right) \, ,
  \\ 
  K_-^{(2)}(H) = \frac{2}{H^2}\left(  1 - \sqrt{1 - H^2} \right) \, .
\end{eqnarray}
\end{subequations}
For all values of $H < 1$, this implies a larger range of unstable
wave numbers as compared to the classical case $H = 0$.   

Let us now assume that the unperturbed Wigner distributions have
Lorentzian form, in analogy to the case of a one-stream plasma, i.e.\ 
\begin{equation}
  W_{j0}(p) = \frac{n_{0j}}{\pi}\frac{p_{Tj}}{(p - p_{0j})^2 +
  p_{Tj}^2} \ ; \ j = 1, 2 \ .
\end{equation}
From Eq.\ (\ref{eq:twostreamdisp}) we then obtain
\begin{eqnarray}
  1 &=& \frac{\omega_{p1}^2}{[\Omega - (p_{01} - ip_{T1})K/m]^2 -
  \hbar^2K^4/4m^2} \nonumber \\
   && + \frac{\omega_{p2}^2}{[\Omega - (p_{02} -
  ip_{T2})K/m]^2 - \hbar^2K^4/4m^2} \ .
\end{eqnarray}

Following Ref.\ \cite{Haas-Manfredi-Feix}, we consider the case when
$p_{01} = -p_{02} \equiv p_0$ and $n_{01} = n_{02} = n_0/2$, 
while for the statistical broadening we assume $p_{T1} = p_{T2} \equiv p_T$.
Using the dimensionless variables given by (\ref{eq:dimlessvar}), we
get 
\begin{equation}\label{eq:special}
  (\bar{\Omega} + i\alpha\bar{K})^2 = 
    \frac{1}{2} + \bar{K}^2 + \frac{H^2\bar{K}^4}{4}
    \pm \frac{1}{2}\sqrt{1 + 8\bar{K}^2 + 4H^2\bar{K}^6} \ , 
\end{equation}
where we have introduced the relative broadening $\alpha \equiv
p_T/p_0$. Thus, in the limit $p_T \rightarrow 0$, we regain the result
of Eq.\ (\ref{eq:constraint2}) and Ref.\ \cite{Haas-Manfredi-Feix}. 
However, in the previously unstable region we now obtain
\begin{eqnarray}
  {\rm Im}(\bar{\Omega}) &=& -\alpha\bar{K} + \left[ \frac12\left( 1 +
  8\bar{K}^2 + 4H^2\bar{K}^6 \right)^{1/2} \right.\nonumber \\
  &&\qquad\qquad\quad \left. - \frac{1}{2} - \bar{K}^2 -
  \frac{H^2\bar{K}^4}{4} \right]^{1/2} \ .
\end{eqnarray}
Again, the broadening $\alpha$ tends to suppress the growth, and the
condition ${\rm Im}(\bar{\Omega}) > 0$ is now given by
\begin{eqnarray}
  \alpha &<& \frac{1}{\bar{K}}%
    \left[ \frac12\left( 1 +
    8\bar{K}^2 + 4H^2\bar{K}^6 \right)^{1/2} \right. \nonumber \\ 
    && \qquad\qquad \left. - \frac12 - \bar{K}^2 - \frac{H^2\bar{K}^4}{4} \right]^{1/2}
    .
\end{eqnarray}

In the classical limit $H \rightarrow 0$, the region of unstable
$\bar{K}$-values is reduced to $\bar{K} < K_c$ by the damping effect,
where 
\begin{equation}
  K_c = \frac{\sqrt{1 - \alpha^2}}{1 + \alpha^2} < 1 \ .
\end{equation}
Clearly, for $\alpha \geq 1$, no instability is possible for any
$\bar{K}$. Another illustration of this is the small-$\bar{K}$
expansion of the growth rate, which reads
\begin{equation}
  {\rm Im}(\bar{\Omega}) \simeq (1 - \alpha)\bar{K}
\end{equation}

The stabilizing influence of $\alpha$ in the general case of $H \neq 
0$ can be inferred as follows: \\
Consider first the case of small $\bar{K}$, while keeping
$H^2\bar{K}^2/4 \sim {\mathscr O}(1)$, i.e.\ we investigate the
growth rate close to the stability boundary. In this limit we obtain
\begin{equation}
  {\rm Im}(\bar{\Omega}) \simeq \left( \sqrt{1 - \frac{H^2\bar{K}^2}{4}}
  - \alpha \right)\bar{K} \ ,
\end{equation}
which clearly shows the stabilizing effect of the damping. In
particular, the stability threshold is now given by
\begin{equation}
  H = \frac{2}{\bar{K}}\sqrt{1 - \alpha^2} \ , 
\end{equation}
Qualitatively this implies a lowering of the upper threshold curve for
small $\bar{K}$ and a concomitant decrease of the region
of instability.

Consider next the limit $\bar{K} \gg 1$, while still assuming
$H^2\bar{K}^2/4 \sim {\mathscr O}(1)$, i.e.\ we examine the effects of
the damping on the narrow instability region, see Fig.\
\ref{fig1}. Introduce the notation
\begin{equation}
  \Delta h \equiv 1 - \frac{H^2\bar{K}^2}{4} \ .
\end{equation}
The growth rate can then be written as
\begin{equation}
  {\rm Im}(\bar{\Omega}) \simeq -\alpha\bar{K} + \sqrt{\Delta h\left(
  \frac{1}{2} - \bar{K}^2\Delta h \right)} \ ,
\end{equation}
and the stability thresholds become determined by
\begin{equation}
  \Delta h = \frac{1}{4\bar{K}^2} \pm \sqrt{\frac{1}{16\bar{K}^4} -
  \alpha^2} \ .
\end{equation}
When $\alpha = 0$, we regain the previous limit curves $\Delta h = 0$
and $\Delta h = 1/(2\bar{K}^2)$. The effect of a nonzero $\alpha$ is
to narrow the instability region and to terminate it at the finite wave
number $\bar{K} = 1/(2\sqrt{\alpha})$. For increasing $\alpha$, the
unstable region decreases and, as in the case of small $\bar{K}$, we
expect the instability to be essentially quenched for $\alpha \gtrsim 1$ .

\section{Discussion}

In this work, we have presented an analysis of a multi-stream quantum
plasma, including the effect of phase noise in the beam wave
function. As compared the fluid description of a quantum plasma used
in Ref.\ \cite{Haas-Manfredi-Feix}, the present analysis is based on
the quantum mechanical Wigner formalism. The phase noise, or partial
incoherence, of the beam wave functions is shown to give rise to a
Landau-like damping effect, which tends to suppress the instabilities
occurring in both the one- and two beam cases. The damping rate increases
with increasing degree of incoherence as expressed by the width of the
probability distribution function for the phase noise. 
The physical origin of this damping
effect is the non-coherent properties of the beam wave function as
opposed to the 
wave-particle interaction characteristic of the conventional Landau
damping. The new Landau-like effect is not a true wave damping, but a
conservative rearrangement of the spectrum of the beam wave function. 
This phenomenon has
recently attracted considerable
interest, both
theoretically\cite{Hall-etal,Christodoulides-etal} 
and experimentally \cite{Mitchell-etal}, 
within the area of nonlinear
optics, where it has been 
shown to suppress the modulational and self-focusing instabilities
\cite{Bang-Edmundson-Krolikowski,Soljacic-etal,Anastassion-etal}, 
e.g.\ for optical beams in nonlinear photo-refractive media. The
present work is the first attempt to extend this theory to a quantum
plasma.



\end{document}